\RequirePackage{fix-cm}
\documentclass[twocolumn]{svjour3}       
\smartqed  
\usepackage{graphicx}
\usepackage{booktabs}
\usepackage{amsmath}

\journalname{myjournal}
\usepackage{ifthen, mciteplus} 

\usepackage{subfig}
\newboolean{uprightparticles}
\setboolean{uprightparticles}{false} 

\usepackage{xspace} 
\usepackage{upgreek}

\def\lhcb   {\mbox{LHCb}\xspace}

\def\lhc    {\mbox{LHC}\xspace}

\def\velo   {VELO\xspace}

\def\MagUp {\mbox{\em Mag\kern -0.05em Up}\xspace}

\def\hltone {HLT1\xspace}
\def\hlttwo {HLT2\xspace}

\ifthenelse{\boolean{uprightparticles}}%
{

 \def\PDelta      {\ensuremath{\Delta}\xspace}                 
 \def\PXi      {\ensuremath{\Xi}\xspace}                 
 \def\PLambda      {\ensuremath{\Lambda}\xspace}                 
 \def\PSigma      {\ensuremath{\Sigma}\xspace}                 
 \def\POmega      {\ensuremath{\Omega}\xspace}                 
 \def\PUpsilon      {\ensuremath{\Upsilon}\xspace}                 
 
 \def\PB      {\ensuremath{\mathrm{B}}\xspace}                 
                  
 \def\PD      {\ensuremath{\mathrm{D}}\xspace}

 \def\PK      {\ensuremath{\mathrm{K}}\xspace}

 \def\Pi      {\ensuremath{\mathrm{i}}\xspace}

}
{

 \mathchardef\PDelta="7101
 \mathchardef\PXi="7104
 \mathchardef\PLambda="7103
 \mathchardef\PSigma="7106
 \mathchardef\POmega="710A
 \mathchardef\PUpsilon="7107
                  
 \def\PB      {\ensuremath{B}\xspace}                 
                  
 \def\PD      {\ensuremath{D}\xspace}

 \def\PK      {\ensuremath{K}\xspace}

 \def\Pi      {\ensuremath{i}\xspace}

}

\makeatletter
\ifcase \@ptsize \relax
  \newcommand{\miniscule}{\@setfontsize\miniscule{4}{5}}
\or
  \newcommand{\miniscule}{\@setfontsize\miniscule{5}{6}}
\or
  \newcommand{\miniscule}{\@setfontsize\miniscule{5}{6}}
\fi
\makeatother

\DeclareRobustCommand{\optbar}[1]{\shortstack{{\miniscule (\rule[.5ex]{1.25em}{.18mm})}
  \\ [-.7ex] $#1$}}

\def\kaon    {{\ensuremath{\PK}}\xspace}
  \def\Kbar    {{\kern 0.2em\overline{\kern -0.2em \PK}{}}\xspace}

\def\KorKbar    {\kern 0.18em\optbar{\kern -0.18em K}{}\xspace}

\def\KS      {{\ensuremath{\kaon^0_{\mathrm{ \scriptscriptstyle S}}}}\xspace}

  \def\Dbar    {{\kern 0.2em\overline{\kern -0.2em \PD}{}}\xspace}

\def\DorDbar    {\kern 0.18em\optbar{\kern -0.18em D}{}\xspace}

\def\Bbar    {{\ensuremath{\kern 0.18em\overline{\kern -0.18em \PB}{}}}\xspace}

\def\BorBbar    {\kern 0.18em\optbar{\kern -0.18em B}{}\xspace}

  \def\Y#1S{\ensuremath{\PUpsilon{(#1S)}}\xspace}

\def\Lz          {{\ensuremath{\PLambda}}\xspace}
\def\Lbar        {{\ensuremath{\kern 0.1em\overline{\kern -0.1em\PLambda}}}\xspace}
\def\LorLbar    {\kern 0.18em\optbar{\kern -0.18em \PLambda}{}\xspace}

\def\AT#1     {\ensuremath{A_{\mathrm{T}}^{#1}}\xspace}           

\def\C#1      {\ensuremath{\mathcal{C}_{#1}}\xspace}                       
\def\Cp#1     {\ensuremath{\mathcal{C}_{#1}^{'}}\xspace}                    
\def\Ceff#1   {\ensuremath{\mathcal{C}_{#1}^{\mathrm{(eff)}}}\xspace}        
\def\Cpeff#1  {\ensuremath{\mathcal{C}_{#1}^{'\mathrm{(eff)}}}\xspace}       
\def\Ope#1    {\ensuremath{\mathcal{O}_{#1}}\xspace}                       
\def\Opep#1   {\ensuremath{\mathcal{O}_{#1}^{'}}\xspace}                    

\newcommand{\tev}{\ifthenelse{\boolean{inbibliography}}{\ensuremath{~T\kern -0.05em eV}}{\ensuremath{\mathrm{\,Te\kern -0.1em V}}}\xspace}
\newcommand{\gev}{\ensuremath{\mathrm{\,Ge\kern -0.1em V}}\xspace}
\newcommand{\mev}{\ensuremath{\mathrm{\,Me\kern -0.1em V}}\xspace}
\newcommand{\kev}{\ensuremath{\mathrm{\,ke\kern -0.1em V}}\xspace}
\newcommand{\ev}{\ensuremath{\mathrm{\,e\kern -0.1em V}}\xspace}
\newcommand{\gevc}{\ensuremath{{\mathrm{\,Ge\kern -0.1em V\!/}c}}\xspace}
\newcommand{\mevc}{\ensuremath{{\mathrm{\,Me\kern -0.1em V\!/}c}}\xspace}
\newcommand{\gevcc}{\ensuremath{{\mathrm{\,Ge\kern -0.1em V\!/}c^2}}\xspace}
\newcommand{\gevgevcccc}{\ensuremath{{\mathrm{\,Ge\kern -0.1em V^2\!/}c^4}}\xspace}
\newcommand{\mevcc}{\ensuremath{{\mathrm{\,Me\kern -0.1em V\!/}c^2}}\xspace}

\def\cm   {\ensuremath{\mathrm{ \,cm}}\xspace}

\def\mm   {\ensuremath{\mathrm{ \,mm}}\xspace}

\def\mhz  {\ensuremath{{\mathrm{ \,MHz}}}\xspace}
\def\khz  {\ensuremath{{\mathrm{ \,kHz}}}\xspace}
\def\hz   {\ensuremath{{\mathrm{ \,Hz}}}\xspace}

\def\gsim{{~\raise.15em\hbox{$>$}\kern-.85em
          \lower.35em\hbox{$\sim$}~}\xspace}
\def\lsim{{~\raise.15em\hbox{$<$}\kern-.85em
          \lower.35em\hbox{$\sim$}~}\xspace}

\def\tell1  {TELL1\xspace}
\def\ukl1   {UKL1\xspace}

\newcommand{\ie}{\mbox{\itshape i.e.}\xspace}

\newboolean{inbibliography}
\setboolean{inbibliography}{true} 
\usepackage{tikz}
\usetikzlibrary{backgrounds,fit}
\usetikzlibrary{shapes.geometric, arrows}

\newboolean{articletitles}
\setboolean{articletitles}{true} 

\usepackage{lineno}

\DeclareGraphicsExtensions{.jpg,.pdf,.png}

\tikzstyle{startstop} = [rectangle, rounded corners, minimum width=3cm, minimum height=1cm, text centered, text width=3cm, draw=black, fill=black!5!white, font=\footnotesize]
\tikzstyle{process} = [rectangle, minimum width=3cm, minimum height=1cm, text centered, text width=3cm, draw=black, fill=black!5!white, font=\footnotesize]
\tikzstyle{decision} = [diamond, aspect=3, minimum width=0.8cm, minimum height=0.3cm, text centered, text width=2.5cm, draw=black, fill=white!90!yellow, font=\footnotesize, yshift=-0.2cm]

\tikzstyle{optional_decision} = [diamond, dashed, aspect=3, minimum width=0.8cm, minimum height=0.3cm, text centered, text width=2cm, draw=black, fill=white!90!yellow, font=\footnotesize, yshift=-0.2cm]
\tikzstyle{process_description} = [rectangle, minimum width=3cm, minimum height=1cm, text centered, text width=3cm, font=\footnotesize]
\tikzstyle{process_cpu} = [rectangle, minimum width=3cm, minimum height=1cm, text centered, text width=3cm, draw=black, fill=white!85!blue, font=\footnotesize]

\tikzstyle{arrow} = [thick,->,>=stealth]
\tikzstyle{dashed_arrow} = [thick,->,>=stealth,dashed]

\begin{document}\sloppy

\title{Hybrid seeding: A standalone track reconstruction algorithm for scintillating fibre tracker at LHCb\thanks{R.~Quagliani acknowledges support of the European Research Council Consolidator grant RECEPT 724777.}\thanks{L.~Henry and S.~Aiola acknowledge support of the European Research Council Consolidator grant SELDOM 771642.}\thanks{C.~Marin Benito acknowledges support of the Agence Nationale de la Recherche grant BACH.}}


\author{S.~Aiola\thanks{*Corresponding authors: Louis Henry louis.henry@cern.ch, Renato Quagliani renato.quagliani@cern.ch \vspace{5mm}}
  \and
  Y.~Amhis \and
  P.~Billoir \and
  B.~Kishor~Jashal \and
  L.~Henry$^*$ \and
  A.~Oyanguren~Campos \and
  C.~Marin Benito \and
  F.~Polci \and
  R.~Quagliani$^*$ \and
  M.~Schiller \and 
  M.~Wang
}

\institute{R.~Quagliani \and F.~Polci \and P.Billoir
\at LPNHE, Sorbonne Universit\'{e}, Paris Diderot Sorbonne Paris Cit\'{e}, CNRS/IN2P3, Paris, France
\and
Y.~Amhis \and C.~Marin Benito \at Universit\'e Paris-Saclay, CNRS/IN2P3, IJCLab, Orsay, France
\and
L.~Henry \at IFIC, Valencia, Spain and Universit\`{a} di Milano, Milano, Italy
\and
A.~Oyanguren Campos \and B.~Kishor~Jashal \at IFIC, Valencia, Spain
\and
S.~Aiola \at Universit\`{a} di Milano, Milano, Italy
\and 
M.~Schiller  \at University of Glasgow, UK
\and
M.~Wang \at Center for High Energy Physics, Tsinghua University, Beijing, China
}

\date{Received: date / Accepted: date}

\maketitle

\newcommand{\HS}{{\tt Hybrid seeding}\xspace}
\newcommand{\scifi}   {SciFi\xspace}
\newcommand{\downtrack}{Downstream\xspace}
\newcommand{\longtrack}{Long\xspace}
\newcommand{\uv}   {\ensuremath{u/v}\xspace}
\newcommand{\xz} {$x$-$z$\xspace}
\newcommand{\zref} {\ensuremath{z_{\rm ref}}\xspace}
\newcommand{\zbar} {\ensuremath{\bar{z}}\xspace}

\begin{abstract}
We describe  the \HS, a stand-alone pattern recognition algorithm aiming at finding charged particle trajectories for the \lhcb upgrade.
A significant improvement to the charged particle reconstruction efficiency is accomplished by 
exploiting the knowledge of the \lhcb magnetic field and the position of energy deposits in 
the scintillating fibre tracker detector. Moreover, we achieve a low fake rate and a
small contribution to the overall timing budget of the \lhcb real-time data processing.

\end{abstract}

\begin{acknowledgements}
We would like to thank A. Hennequin for many fruitful discussions. We also thank the LHCb RTA team for supporting this publication and reviewing this work. We thank the technical and administrative staff at the LHCb institutes.
\end{acknowledgements}

\section{Introduction}
\label{intro}
The LHCb detector~\cite{LHCb-DP-2008-001} is undergoing a major upgrade in preparation of the Run~3 data taking at the LHC, starting in 2021~\cite{LHCb-TDR-012}. 
The expected delivered instantaneous luminosity is ${\cal L}= 2\times 10^{33}\cm^{2}s^{-1}$, corresponding to an average of seven proton-proton interactions per bunch collision. 

The entire charged particle reconstruction (tracking) system of the \lhcb detector is renewed as part of this upgrade. In particular, the tracker placed downstream of the \lhcb dipole magnet is replaced by a scintillating fibre tracker (\scifi) described in detail in Ref.~\cite{LHCb-TDR-015}. 
The \scifi consists of three stations (T1, T2, T3), each composed of four layers of stacked scintillating fibres. The layers within one station are separated from each other by an air-filled gap of 50 mm and they are oriented in a stereo configuration ($x$-$u$-$v$-$x$). 
For the sake of mechanical stability, the scintillating fibres in the  $x$-layers are  strictly vertical, so that they have a slight tilt with respect to the $y$ axis, which is perpendicular to the beam axis in the usual \lhcb coordinate system.
The $u/v$ layers are rotated in the $x-y$ plane by the stereo angle, $\alpha$, equal to $+5^{\circ}$ and $-5^{\circ}$ for the $u$ and $v$ layers, respectively. 

An algorithm relying solely on the information provided by this tracker, called \HS, is described in this paper. This algorithm allows an efficient reconstruction of tracks from particles with momenta down to 1.5 \gevc. The track segments reconstructed by this algorithm are used as seed for other pattern recognition algorithms in LHCb.

The \lhcb upgrade trigger strategy relies on two software stages called \hltone and \hlttwo~\cite{LHCb-TDR-016}.
The \hltone stage performs a partial reconstruction of the event in order to select general heavy-flavour physics signatures, such as tracks from displaced vertices, tracks with high transverse momentum or muons carrying large transverse momentum.
The \hlttwo reconstruction exploits the reduced event rate after the \hltone selections to perform a full reconstruction of the event, adding offline-quality particle identification information and aiming at reconstructing all the tracks in the event. This reconstruction also benefits from the real-time alignment and calibration procedure developed for the Run~2 data taking~\cite{LHCb-DP-2019-001}.
Within this scheme, and in its current version, the \HS is designed to be executed within \hlttwo.
The current goal is to run the \hlttwo stage at a frequency of 1\mhz using around 1000 CPU nodes~\cite{LHCb-TDR-016}.

\section{Motivation}
\label{sec:motivation}

\begin{figure}
    \centering
    \includegraphics [scale =0.30
    ]{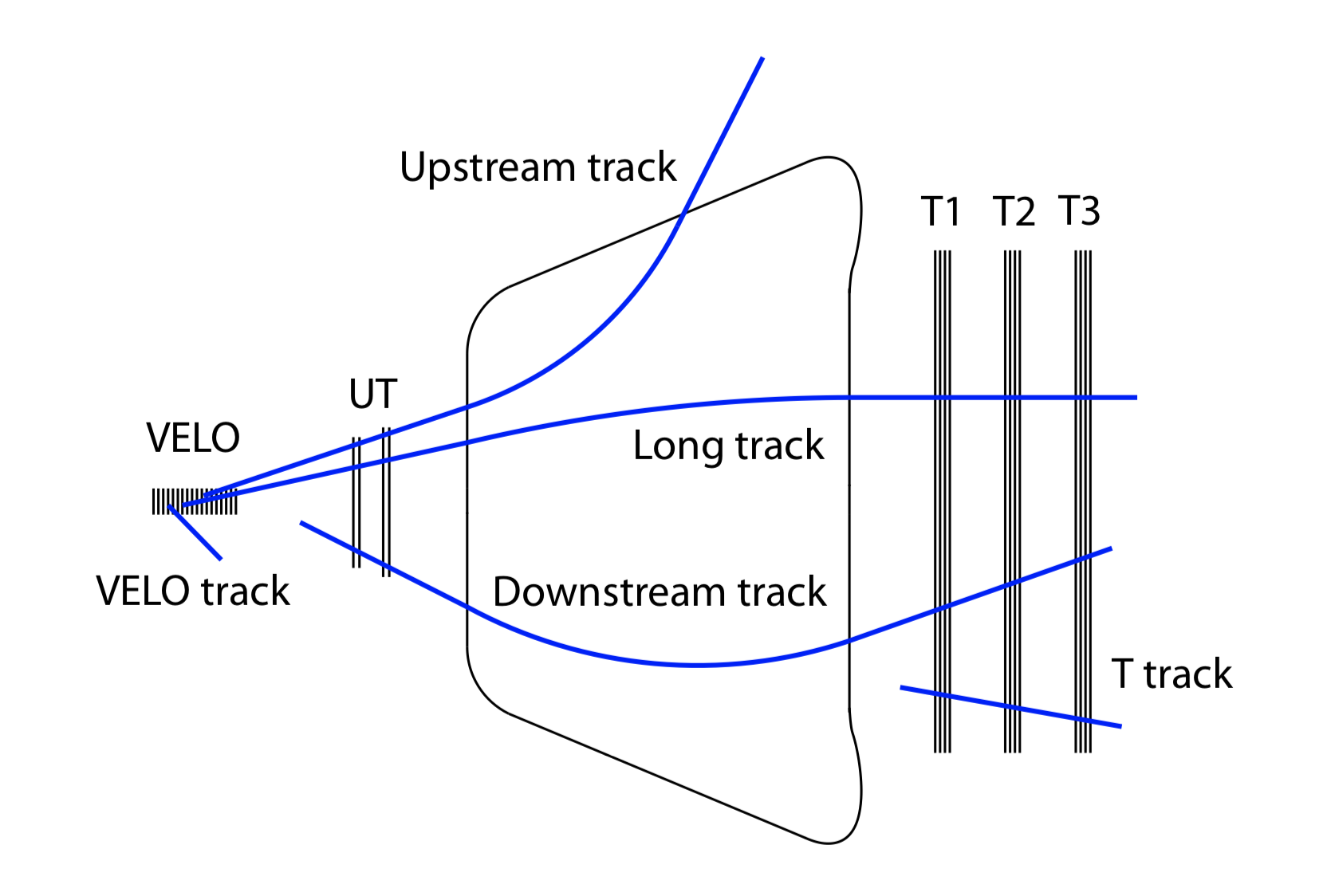} 
    \caption{Track types defined in \lhcb. It is worth noting that Long tracks do not need to leave hits in the Upstream Tracker (UT). The SciFi is composed of the T1, T2 and T3 stations, and the \lhcb magnet is represented between the UT and the SciFi. Upstream tracks are formed by a combination of hits in the \velo and the UT.}
    \label{fig:lhcb}
\end{figure}

The \lhcb detector at the \lhc is specialised in the study of heavy hadrons.
Physics analyses in \lhcb rely mainly on two track types, called ``\longtrack'' and ``\downtrack'' as defined in Fig.~\ref{fig:lhcb}. \longtrack tracks are reconstructed using energy deposits (hits)  from at least the vertex locator (\velo)~\cite{LHCb-TDR-013} and the \scifi sub-detectors, and represent the majority of tracks used in \lhcb analyses. \downtrack tracks, which are reconstructed using hits in the upstream tracker (UT) and \scifi, typically correspond to the decay products of long-lived hadrons such as \KS mesons and \Lz baryons.

The \lhcb track reconstruction~\cite{LHCb-DP-2013-002} consists of two main consecutive steps. First, the pattern recognition algorithms create track segments by connecting hits from the sub-detectors: the \velo, the UT and the \scifi. Second, the track fitting provides the track candidates, with a track quality expressed in terms of a $\chi^{2}$ per degrees of freedom.

The \HS~\cite{Quagliani:2296404} is a stand-alone algorithm, \ie it does not rely on any input from other algorithms in the tracking sequence. It produces track segments inside of the \scifi, used later by other pattern recognition algorithms. In addition to allowing the reconstruction of \downtrack tracks, these segments can be combined with track segments in the \velo to form \longtrack tracks.

\begin{figure}
    \centering
    \includegraphics[width = 0.49\textwidth]{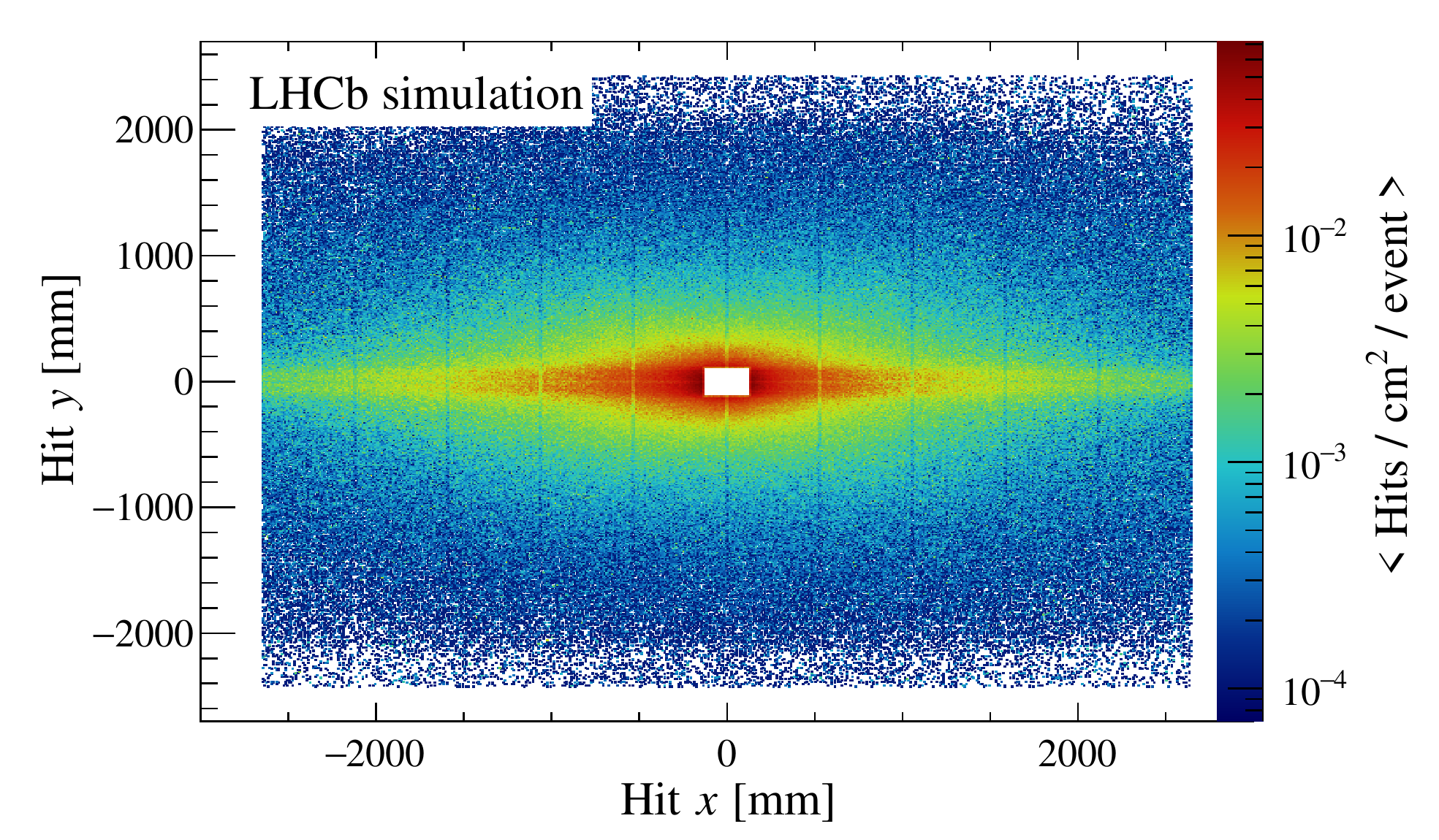}
    \includegraphics[width = 0.49\textwidth]{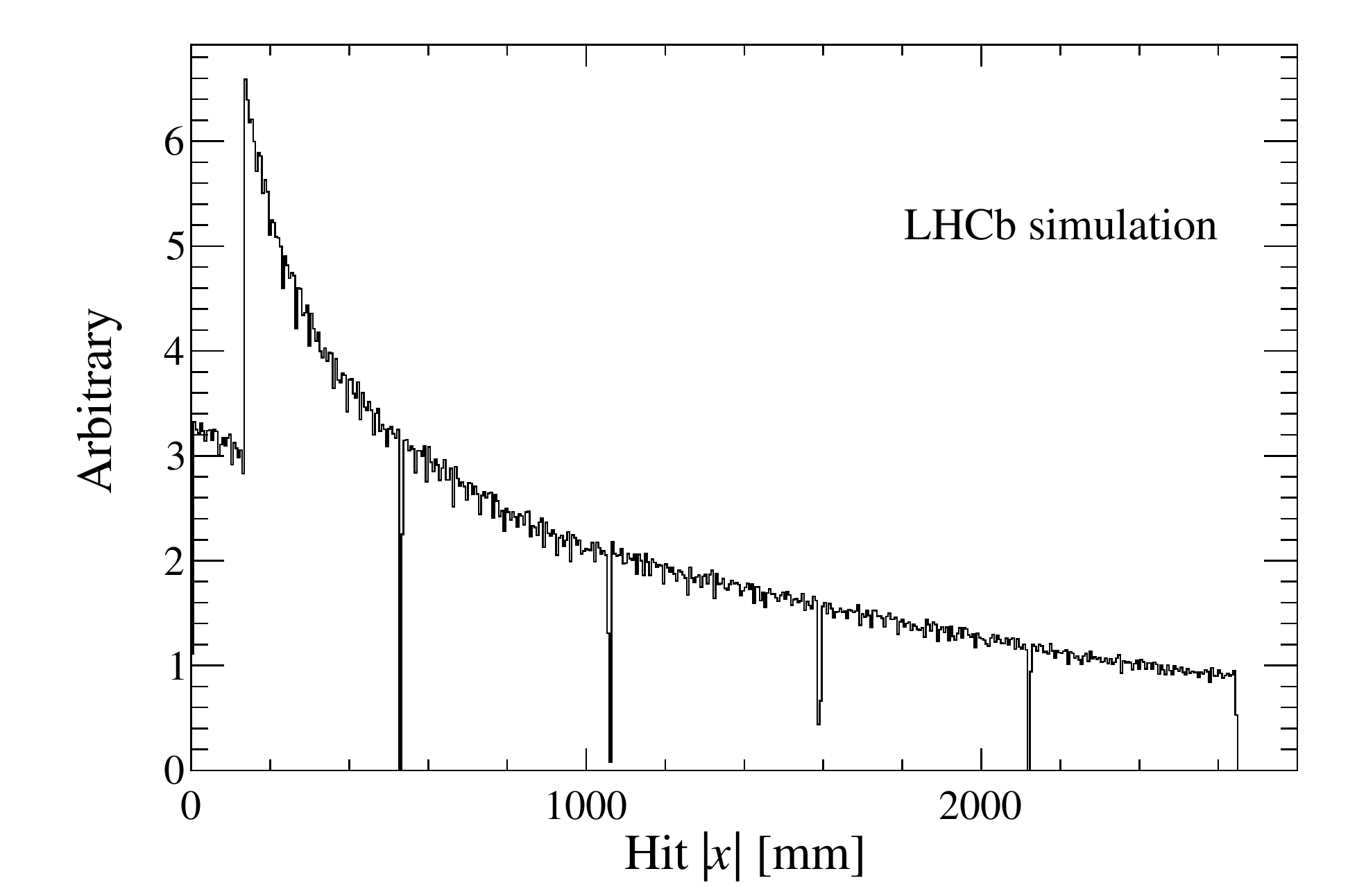}
    \caption{Top: event average hit density in a single layer of the \scifi. The hole in the middle corresponds to the beam hole. Bottom: information provided after detector readout to the \HS ($y$-information is integrated out).
    }
    \label{fig:occupancy}
\end{figure}

The average occupancy expected in the \scifi in Run 3, shown in Fig.~\ref{fig:occupancy}, sets a strong challenge for the design of the \HS: the high hit multiplicity in the central region makes it difficult to identify the correct hits belonging to a given track path. The non-negligible residual magnetic field in the \scifi region further complicates this task, inducing a non-uniform bending of the particle trajectories within the \scifi acceptance. The \HS has been designed to handle  these conditions by exploiting a dedicated parameterisation of the tracks. It provides a high reconstruction efficiency while maintaining a low rate of fake track candidates.  

\section{The algorithm}

\label{sec:algorithm}
The \scifi detector provides measurements of hit positions as $(x,z)$ coordinates, which allow a direct extraction of track patterns in the bending plane of the \lhcb detector (along the $x$ direction). The track patterns in the non-bending plane (along the $y$ direction) are measured combining information from the \xz
patterns with hits from the \uv layers.
For this reason, the algorithm is designed to first build the projection of the track candidates in the \xz plane, called {\it seed} candidates, by combining measurements from the six $x$-layers only. 
Later on, a search for matching hits in the non-bending plane is performed using the information from \uv-layers. The $y$-$z$ patterns are built through the relation 
$x^{\uv}_{\rm measured}-x_{z} = y_{z}\times \tan(\alpha)$, where $x_z$ and $y_z$ define the true position of a track in the detection layer placed at position $z$ and $\alpha$ is the stereo angle defined in Sec.~\ref{intro}. These two pieces of information can be constrained thanks to a dedicated track model parameterisation.
The final output of the \HS is a 3-dimensional track segment that can be combined with information from other detectors. 

\begin{figure}
    \centering
    \includegraphics[width = 0.49\textwidth]{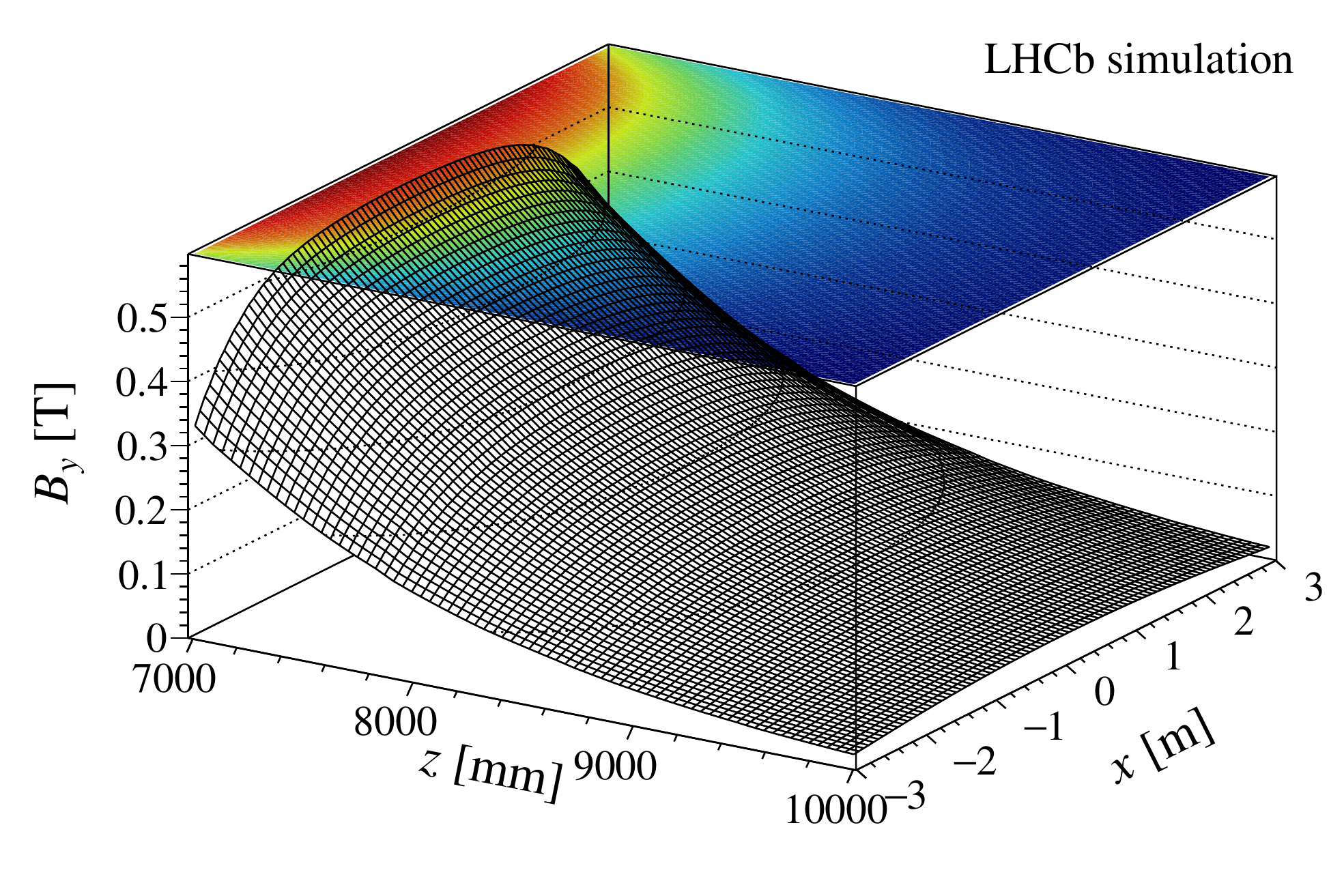}
    \caption{Dependency of the $B_{y}$  magnetic field component as a function of the bending plane in the acceptance region of the SciFi detector. The magnetic field intensity decreases as a function of $z$ and can be written in a first approximation as $B_{y}=B_{0}+B_{1}\cdot z $, where $B_{1}/B_{0}$ is roughly constant.}
    \label{fig:scifimagfield}
\end{figure}

The first task of the \HS is to reduce the number of combinations when finding the hits corresponding to a particle's trajectory. 
The trajectory of a track with momentum $\overrightarrow{p}$ is modelled through a double integration of the equation of motion $d\overrightarrow{p}/dt = q \overrightarrow{v}\times \overrightarrow{B} $, for a particle with charge $q$ and velocity $\overrightarrow{v}$.
The residual magnetic field $\overrightarrow{B}$ in the \scifi geometrical acceptance can be in first approximation described by $B_{x}\sim 0$, $B_{y} = B_{0} + B_{1} \cdot \zbar $,  $B_{z}\sim 0$,
where $\zbar = z - \zref$ with $\zref = 8525\mm$. 
In the small track angle approximation 
\begin{align}
    x(z) &= a_x + b_x\zbar +c_x\zbar^2\left(1+d_x\zbar\right)
\label{eq:trackmodel_x}
    \\
    y(z) &= a_y + b_y\zbar,
\label{eq:trackmodel_y}
\end{align}
where $a_x,b_x,a_y,b_y$ correspond respectively to the usual parameters $x,t_x = dx/dz,y,t_y = dy/dz$ at $\zbar$. The $a_x, b_x, c_x, a_y$ and $b_y$ parameters are left free in the track fit. The $d_x$ parameter, which is equal in this approximation to  $B_1/3B_0$, is taken as a constant, estimated from studies on simulated samples. This parameterisation is motivated by the observed variation of the main magnetic field component ($B_{y}$) in the \lhcb bending plane, as shown in Figure~\ref{fig:scifimagfield}. 

A path to collect hits is defined using search windows
assuming that the tracks originate from $(0,0,0)$, which is located inside of the \velo. This hypothesis holds well for all tracks that originate from before the UT.
The number of hit combinations to be processed increases with the size of the search window. Therefore, a progressive ``cleaning" of the tracking environment is performed in 3 iterations. This allows to open a wide search windows in the last iteration, as needed for the reconstruction of low-momentum particles.
This progressive approach is necessary to respect the strict timing constraints of real-time analysis~\cite{LHCb-TDR-016} and to keep the rate of fake tracks low.
Furthermore, each iteration starts with different pairs of $x$-layers, one in T1 and one in T3, in order to compensate for hit detection inefficiency, due to either the fibres themselves ($1-3\,\%$ depending on irradiation~\cite{LHCb-TDR-015}) or inactive zones between fibre mats.
Table~\ref{tab:cases} gives the layers used in each iteration along with their momentum range. The indices "x1" and "x2" designate the first and second $x$-layer of a given station, respectively.
\begin{table}[]
    \centering
    \caption{Definitions of the iterations in the  \lhcb implementation. All search windows are tuned according to simulated tracks that satisfy the momentum condition.}
    \label{tab:cases}
    \begin{tabular}{c|c|c|c}
    \toprule
    \midrule
         Iteration & 1 & 2 & 3 \\
         \midrule
         First $x$-layer & T1x1 & T1x2 & T1x1 \\
         Second $x$-layer & T3x2 & T3x1 & T3x1 \\ 
         Minimum momentum & 5\gevc & 2\gevc & 1.5\gevc \\
         \midrule
         \bottomrule
    \end{tabular}
\end{table}

\begin{figure}
    \centering
    \includegraphics[width = 0.49\textwidth]{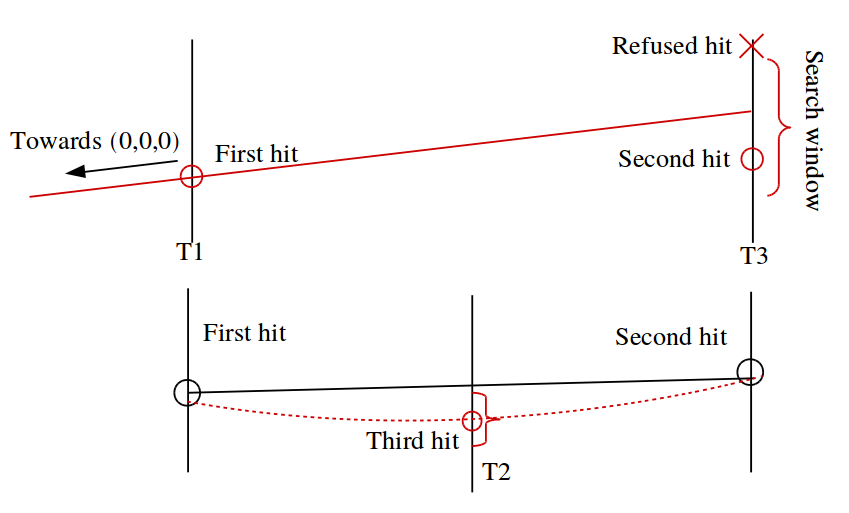}
        \caption{Illustration of the two- and three-hit searches in the $x$-$z$ plane.}
    \label{fig:xz}
\end{figure}

For each hit in the first layer, a search window is opened
in the second layer, according to the considered momentum range, expected charge disparity in different halves of the detector due to the magnetic field, and assuming that the tracks originate from $(0,0,0)$, as illustrated in Figure~\ref{fig:xz}.
The windows are kept wide enough so that the efficiency is large for all particles created before the magnet and for electrons, which can lose energy due to the bremsstrahlung effect~\cite{LHCb-DP-2019-003}.
For each pair of hits, together with the assumption that the track comes from the origin, a momentum hypothesis is derived and used to predict the region in each of the $x$-layers of the T2 station where a hit from the corresponding track is expected.
Once a third hit is found in T2, a curve is defined and remaining hits from leftover $x$-layers are searched for along it to form a track candidate.
Only track candidates with at least five hits, each in a different $x$-layer, are selected.
A fit is then performed according to
Eq.~\ref{eq:trackmodel_x} and track candidates are filtered using a $\chi^2$ per degree of freedom criterion. In this process, outlier hits can be removed, down to a minimum of four hits. 

Hits processed in this  phase of the \HS are sorted by layer and increasing $x$ coordinate.
For a given momentum hypothesis and pair of seed layers, the boundaries of the two-hit search windows are monotonously increasing in $x$. As a result, the search for the second hit is significantly sped up by caching search bounds.
This approach is replicated for the search of the third hit, and of remaining hits, as two-hit combinations corresponding to the same first hit are naturally sorted by slope.
Finally, the position of the first hit in T2 that matches a two-hit combination built from a given hit in T1 is saved.  The next search for the third hit starts from the previously formed two-hit combination.
This optimisation allows to move the search window in each T2 layer by only a few hits when iterating over T1 hits.

Only about a third of \xz candidates are real tracks, the rest are fake combinations. Using $x(z)$ from Eq.~\ref{eq:trackmodel_x}, acceptance windows are defined in the six \uv layers of the SciFi, and hits are collected. It is worth noting that, due to the arrangement of fiber mats in the \uv layers, tracks that belong to the $y>0$ plane could have hits in the lower part of the layer, and conversely. This is accounted for by considering both halves of the detector in the hit collection part.
Each of the collected hits, along with the $x(z)$ track equation, corresponds to a possible $y$ measurement, as well as a track slope $t_y(z_0)=y/(z_{\rm layer}-z_0)$ measurement. In a
dipolar magnetic field along the $y$ direction, tracks coming from a given $(0,0,z_0)$ point with low values of $t_x$ and $t_y$ are expected to have a quasi constant $t_y(z_0)$.
The current implementation of the \HS is tuned towards the reconstruction of tracks coming from the interaction region, so it only performs
searches for tracks with constant $t_y(z_0) \simeq t_y(0)$.
Although the optimisation is driven by \longtrack tracks, \downtrack tracks and electrons are also reconstructed, thanks to the size of the search windows. 

The search for $t_y(0)$ clusters is performed using a binned implementation of the ``Hough cluster search"~\cite{hough1962method}, with one histogram per \uv-layer. Each bin is set to zero if no corresponding $t_y(0)$ value is found, and one otherwise. An additional attribute of the Hough cluster search algorithm stores the address of hits corresponding to filled bins.
Once the histograms are filled, looking for an actual cluster amounts to summing the values of the six histograms, bin-per-bin, and looking for a bin with a value four or higher, the actual criterion depending on parameters of the algorithm.
When several clusters are found, typically in events with high multiplicity and in busy regions of the detector, only the three largest ones are kept.
A fit is performed on each of the retained clusters, and the hits matching the best linear fit are added to the \xz candidate.
It should be noted that the $y$-$z$ fit model does not enforce any constraint on the origin of the track, thus downstream tracks are not rejected by this step.
The total candidate track is fitted using the full track model defined in Eq.~\ref{eq:trackmodel_x} and filtered using additional $\chi^2$ criteria.
Finally, an additional criterion based the track $\chi^2$ and the number of hits on the segment is applied to tracks in order to determine whether to flag hits belonging to them.
Hits that are flagged to belong to high-quality tracks are not considered in further iterations of the search.

After the previous process has been repeated few times, to cover the whole desired momentum range, a recovery routine is employed to recycle \xz projections for which a final track candidate could not be built by adding \uv hits.
The recycled \xz projections are required to share no hit with any confirmed candidate, in order to reprocess only \xz projections that are likely to be associated to a new track candidate. When those candidates are recovered, the \uv information is added in the same manner as explained earlier but with enlarged search windows for the track criteria selections.

Finally, a clone removal procedure is applied, checking for closest approach of different tracks in T1/2/3 and the number of shared hits. In the case where two tracks are determined to be clones of each other ($>70\%$ shared hits on a given track), the one with the largest number of hits is retained and, in case of equality, the one that has the lower $\chi^2$ from the fit using the full track model is retained.

\section{Performance}
\label{performance}
As detailed in Section~\ref{sec:motivation}, the aim of the \HS algorithm is to reconstruct tracks with very different kinematical and topological properties with the highest possible efficiency.
Table~\ref{tab:performances:efficiency} displays the efficiencies of the \HS over different types of tracks, as calculated using samples of 1000 simulated events. Tracks are considered as reconstructible in the \scifi (``Has T'') if they leave a hit in at least one layer in each station. In Table~\ref{tab:fakerate}, the  evolution of the fake rate is investigated for various simulated events.

The simulated physics samples are selected to reflect the ambitious LHCb physics program. 
The efficiency is close to or larger than 90\,\% for all types of tracks that are relevant to LHCb physics analyses, i.e.  \longtrack and \downtrack tracks.
The inefficiency is due to selection criteria and to the hypotheses made in the \HS algorithm, as described in Sec.~\ref{sec:algorithm}, which 
favour tracks from primary interactions or short-lived decays.
Furthermore, almost half of the tracks originating
from secondary interactions with the detector are reconstructed, which allows to match them with the corresponding energy deposits in the calorimeter in order to tag the calorimeter clusters as ``charged".

\begin{table}[]
    \centering
    \caption{Efficiencies ($\epsilon$) for different kinds of tracks, as obtained from simulated samples of 1000 events.
    The "Has T" requirement states that a given track has at least one associated hit in each T station, and would then be reconstructible by an ideal algorithm.
    Long-lived is defined as having a segment in the UT and T stations, but not in the \velo.
    The "h.i." and "p.p." labels refer to "hadronic interaction" and "pair production", respectively, which are the two categories of material interaction creating charged tracks in the detector.
    If not specified otherwise, the tracks are created through decay of particles originated in a primary vertex.
    }
    \label{tab:performances:efficiency}
    \begin{tabular}{c | c |c}
        \toprule
        \midrule
        Sample & Track type & $\epsilon$ [\%] \\
        \midrule
        &Not\textit{ $e^{\pm}$} \\
         $B_{s} \rightarrow \phi\phi$ & Has T&  82.0\\
         $B_{s} \rightarrow \phi\phi$ & Has T, from h.i. & 43.3\\
        &Not\textit{ $e^{\pm}$, $2<\eta<5$} \\
         $B_{s} \rightarrow \phi\phi$ & Long&  92.4\\
         $B_{s} \rightarrow \phi\phi$ & Long, $p > 5\gevc$& 95.4 \\
         $D^{*} \rightarrow (D\rightarrow K\pi) \pi$ & Long, from D & 93.5\\
         $D^{*} \rightarrow (D\rightarrow K\pi) \pi$ & Long, from D, $p > 5\gevc$& 96.4\\
         $D^+ \rightarrow K^0_{\rm S}\pi^+$ & Long-lived&  90.0\\
         $D^+ \rightarrow K^0_{\rm S}\pi^+$ & Long-lived, $p > 5\gevc$& 95.2\\
         $B_{s} \rightarrow \phi\phi$ & Long, from h.i. & 87.4\\
         \midrule
         &\textit{$e^{\pm}$, $2<\eta<5$} \\
         $B \rightarrow K^{*0}e^{+}e^{-}$ & Long & 89.4\\
         $B \rightarrow K^{*0}e^{+}e^{-}$ & Long, $p > 5\gevc$ & 90.7\\
         $B_{s} \rightarrow \phi\phi$ & Has T, from p.p. & 49.6\\
         $B_{s} \rightarrow \phi\phi$ & Long, from p.p. & 85.0\\
       \midrule
       \bottomrule
    \end{tabular}
\end{table}

\begin{table}[]
    \centering
    \caption{Fake rate and average fake rate as function of the simulated decay, calculated using 1000 events of each. The fake rate is defined as $\sum_{\textrm{event}} n_{\textrm {fake}}/\sum_{\textrm{event}} n_{\textrm{ reconstructed}}$, while the average fake rate is defined as $1/N_{
    \textrm{events}} \cdot \sum_{\textrm{event}} \left(n_{\textrm{fake}}/n_{\textrm{reconstructed}}\right)$.}
    \label{tab:performances:ghostrate}
    \begin{tabular}{c|c c}
        \toprule
        \midrule
         Simulated decay & Fake rate [\%] & Average [\%]\\
		 $B \rightarrow K^{*0}e^{+}e^{-}$ & 9.9 & 6.5\\
		 $B_{s} \rightarrow \phi\phi$ & 11.7 & 6.8\\
		 $D^{*+} \rightarrow (D^{0} \rightarrow K^{-}\pi^{+}) \pi^{+}$ & 10.7 & 6.5\\
		 $Z \rightarrow \mu^{+} \mu^{-}$ & 14.7 & 8.1\\
		 Minimum bias & 7.7 & 4.9\\
       \midrule
       \bottomrule
    \end{tabular}
    \label{tab:fakerate}
\end{table}

\begin{figure}
    \centering
    \includegraphics[width=0.45\textwidth]{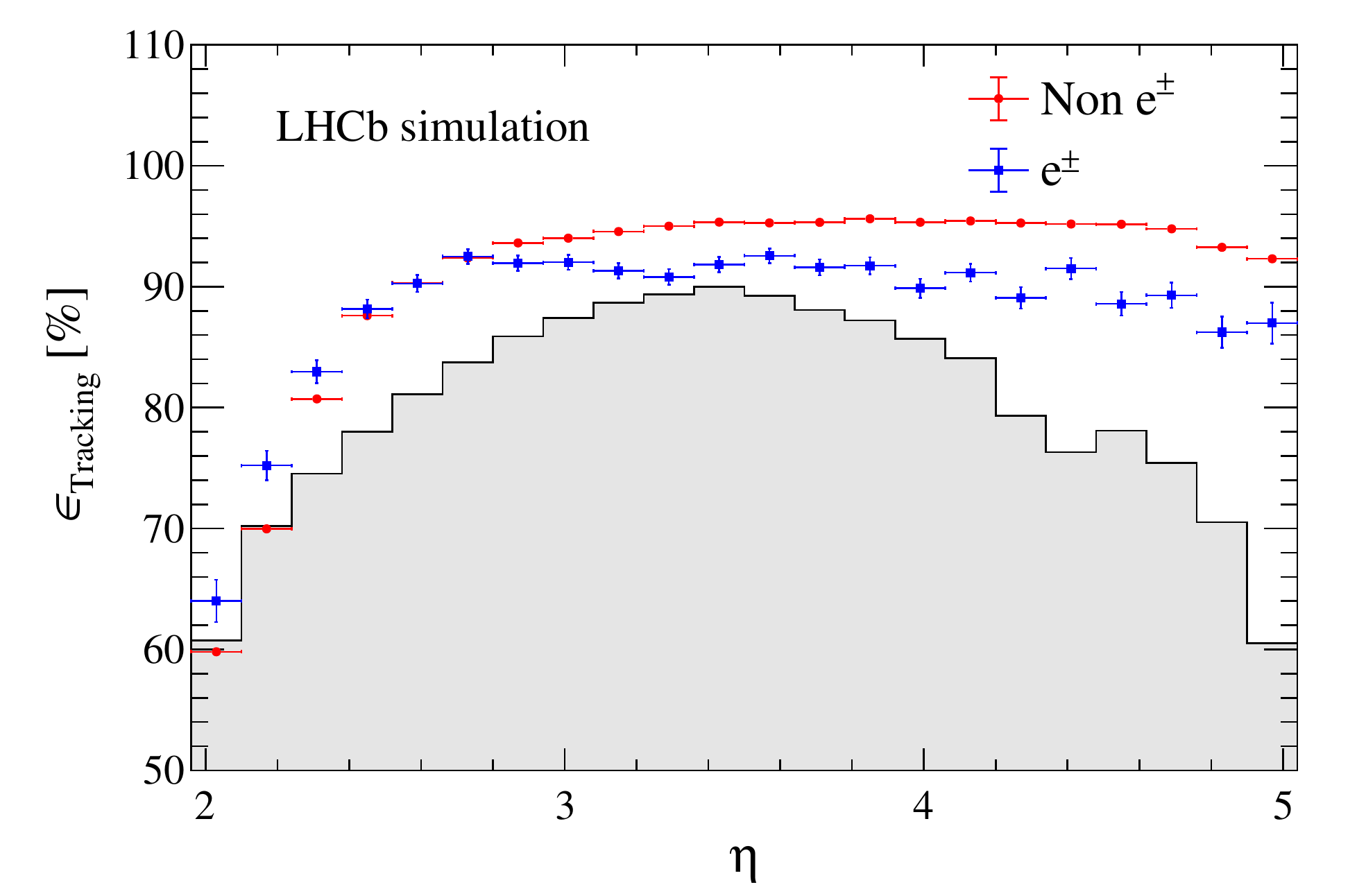}
    \includegraphics[width=0.45\textwidth]{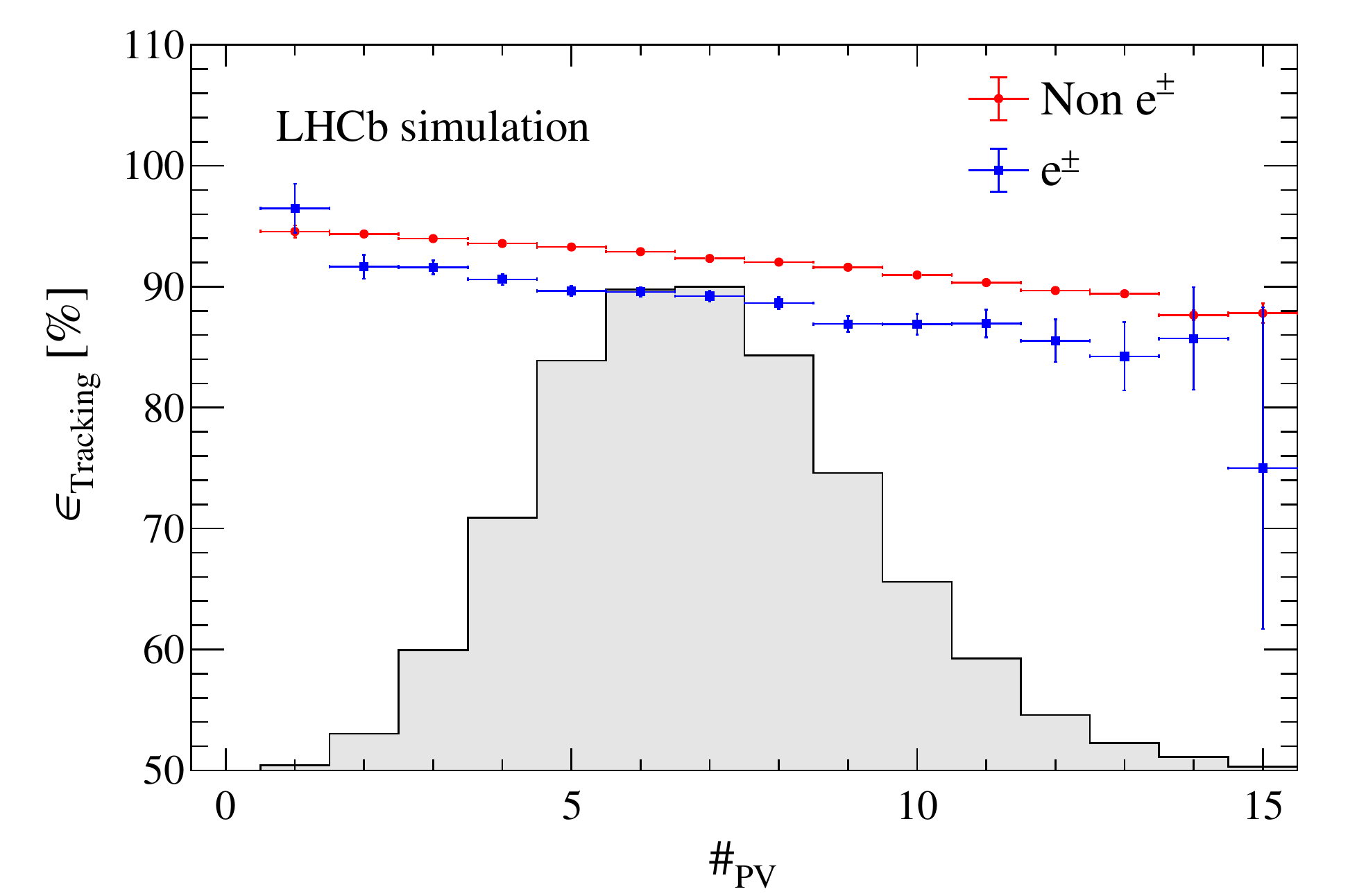}
    \includegraphics[width=0.45\textwidth]{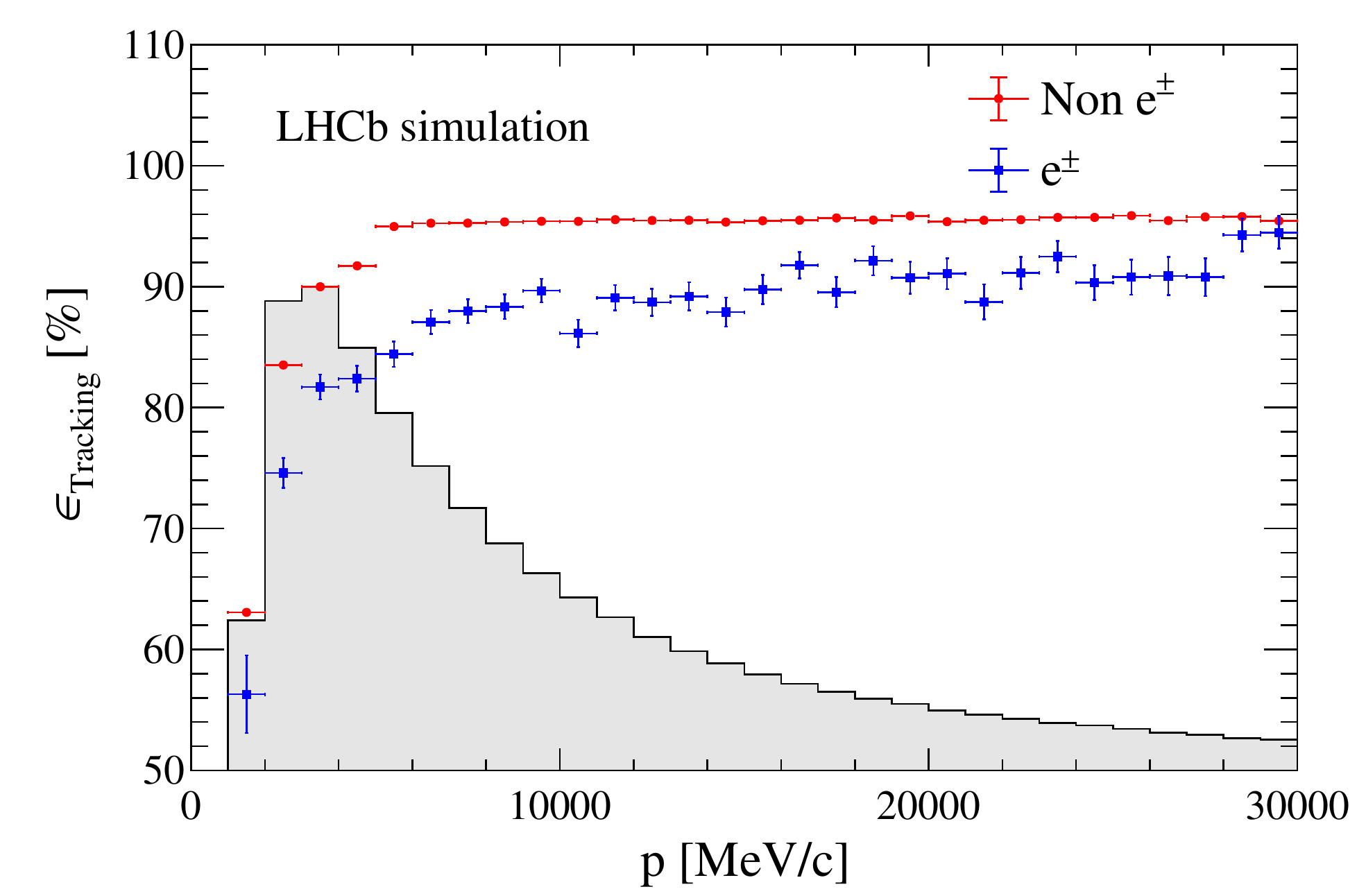}    
    \caption{Dependency of the seeding efficiencies with respect to (top) pseudorapidity, (middle) the number of primary vertices and (bottom) momentum. Blue and red refer to electron and non-electron tracks, respectively. Tracks are taken within the $2 < \eta < 5$ interval.}
    \label{fig:performances:efficiency}
\end{figure}

\begin{figure}
    \centering
    \includegraphics[width=0.45\textwidth]{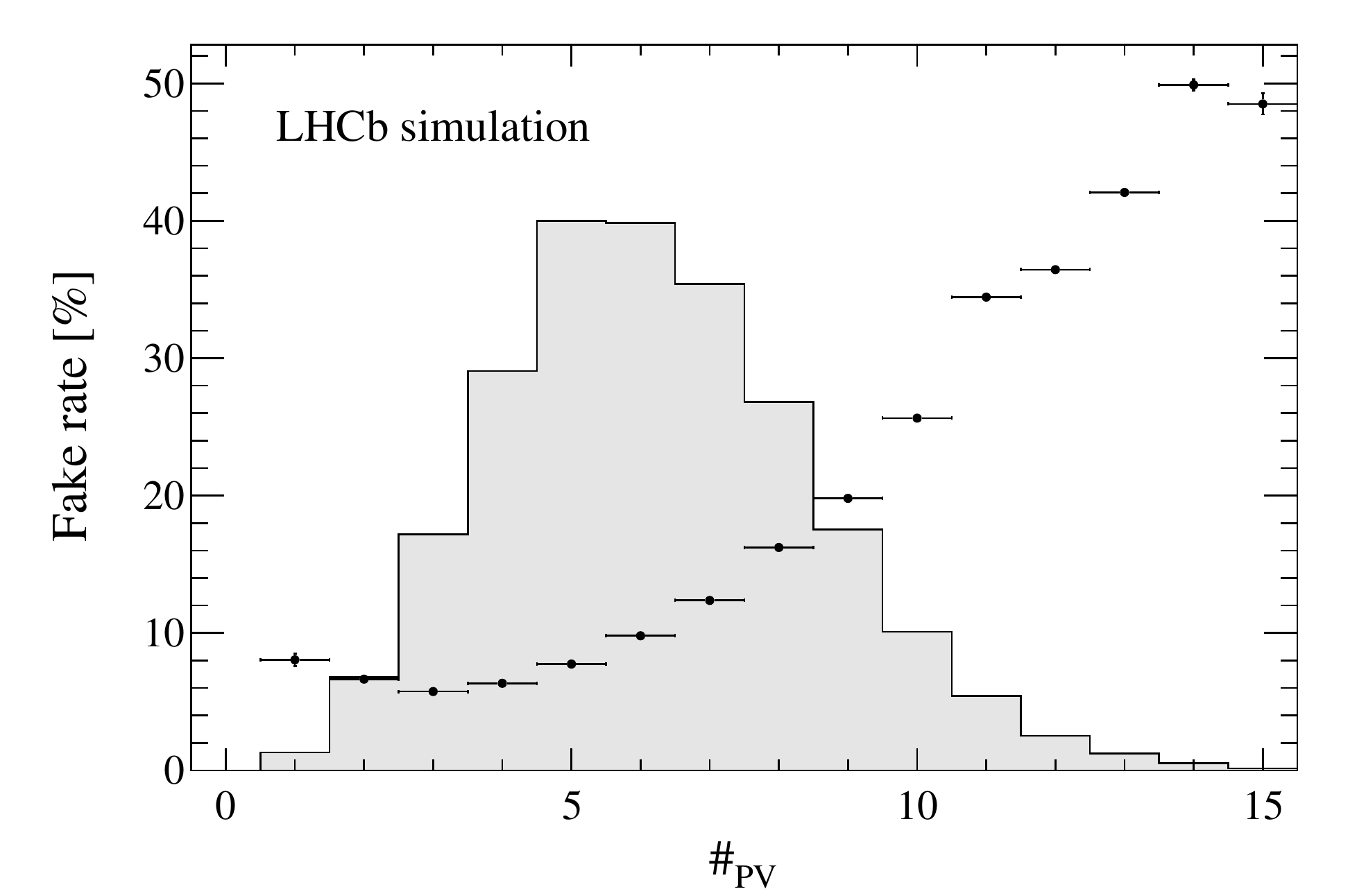}
    \caption{Dependency of the seeding fake rates as a function of the number of primary vertices in the event.}
    \label{fig:performances:fakerate}
\end{figure}

Figure~\ref{fig:performances:efficiency} shows the dependence of the efficiency on the pseudorapidity and the momentum of the tracks, as well as the number of primary vertices.
Efficiency saturates at around 95\,\% for non-electron, high-momentum tracks, and decreases slightly with increasing number of primary vertices.
This is related to the increasing complexity of the underlying event.
Additionally, on Fig.~\ref{fig:performances:fakerate} the dependence of the fake rate on the number of primary vertices is shown, demonstrating an increasing trend with more primary vertices, as expected. 
The larger fake rate at small number of primary vertices can be attributed to an increased proportion of noise and hits from other bunch-crossings. 

Table~\ref{tab:performances:timings} shows the share of the \hlttwo timing dedicated to the \HS algorithm in September 2019 and June 2020, along with the total throughput of the \hlttwo sequence on a reference dual Intel-Xeon-CPU-E5-2630-v4 node.
Using a naive extrapolation by dividing the total throughput by the timing share, the estimated throughput of the seeding sequence has then jumped from around 800\hz/node to around 3.4\khz/node.
This increase in speed has been made possible through many improvements, the most relevant being
\begin{itemize}
    \item a modernisation of the \texttt{C++} code;
    \item a better hit caching and exploitation of the inherent ordering of hit containers;
    \item the replacement of an unbinned Hough cluster approach by the current binned one, which preserves ordering and layer information;
    \item the pre-caching of topological information for a fast parabola and tolerance window calculation.
\end{itemize}

\begin{table}[]
    \centering
    \caption{Overall throughput of the \hlttwo sequence, as well as the timing share dedicated to the seeding and an estimate of the seeding-only throughput.}
    \begin{tabular}{c c c}
    \toprule
    \midrule
          & Sept. 2019 & June 2020 \\
          \midrule
          \hlttwo throughput [\hz] & 90 & 122.5 \\
          Seeding share [\%] & 11 & 3.5 \\
          Seeding throughput [\hz] & 818 & 3441\\
\midrule
\bottomrule
    \end{tabular}
    \label{tab:performances:timings}
\end{table}


Finally, the timing breakdown and physics performance of the \HS is comparable to an alternative approach that builds \longtrack tracks starting from \velo segments instead~\cite{LHCb-TDR-015}.
This ensures that the \HS can play a complementary role in the reconstruction of \longtrack tracks, which form the core of the \lhcb physics program.

\section{Conclusion}
\label{sec:conclusion} 
We have presented an implementation of the \HS, a pattern recognition algorithm designed to reconstruct track segments using information from the \scifi tracker installed for the \lhcb upgrade.
The use of an iterative approach allows to handle the high multiplicity of hits in the detector, while keeping high efficiencies on a variety of tracks and a low fake rate.
Performance studies show that this algorithm improves the reconstruction efficiency for a wide range of particles  with respect to the former baseline presented in Ref.~\cite{LHCb-TDR-015}. It is currently compatible with the timing and efficiency constraints of real-time analysis~\cite{LHCb-TDR-021}, and therefore belongs to the baseline scenario of the \lhcb trigger in Run~3.

%
%

\bibliographystyle{LHCb}
\bibliography{main,LHCb-DP,LHCb-TDR}   

\ifx\mcitethebibliography\mciteundefinedmacro
\PackageError{LHCb.bst}{mciteplus.sty has not been loaded}
{This bibstyle requires the use of the mciteplus package.}\fi
\providecommand{\href}[2]{#2}
\begin{mcitethebibliography}{10}
\mciteSetBstSublistMode{n}
\mciteSetBstMaxWidthForm{subitem}{\alph{mcitesubitemcount})}
\mciteSetBstSublistLabelBeginEnd{\mcitemaxwidthsubitemform\space}
{\relax}{\relax}

\bibitem{LHCb-DP-2008-001}
LHCb collaboration, A.~A. Alves~Jr.\ {\em et~al.},
  \ifthenelse{\boolean{articletitles}}{\emph{{The \lhcb detector at the LHC}},
  }{}\href{https://doi.org/10.1088/1748-0221/3/08/S08005}{JINST \textbf{3}
  (2008) S08005}\relax
\mciteBstWouldAddEndPuncttrue
\mciteSetBstMidEndSepPunct{\mcitedefaultmidpunct}
{\mcitedefaultendpunct}{\mcitedefaultseppunct}\relax
\EndOfBibitem
\bibitem{LHCb-TDR-012}
{LHCb collaboration}, \ifthenelse{\boolean{articletitles}}{\emph{{Framework TDR
  for the LHCb Upgrade: Technical Design Report}}, }{}
  \href{http://cdsweb.cern.ch/search?p=CERN-LHCC-2012-007&f=reportnumber&action_search=Search&c=LHCb+Reports}
  {CERN-LHCC-2012-007}, 2012\relax
\mciteBstWouldAddEndPuncttrue
\mciteSetBstMidEndSepPunct{\mcitedefaultmidpunct}
{\mcitedefaultendpunct}{\mcitedefaultseppunct}\relax
\EndOfBibitem
\bibitem{LHCb-TDR-015}
{LHCb collaboration}, \ifthenelse{\boolean{articletitles}}{\emph{{LHCb Tracker
  Upgrade Technical Design Report}}, }{}
  \href{http://cdsweb.cern.ch/search?p=CERN-LHCC-2014-001&f=reportnumber&action_search=Search&c=LHCb+Reports}
  {CERN-LHCC-2014-001}, 2014\relax
\mciteBstWouldAddEndPuncttrue
\mciteSetBstMidEndSepPunct{\mcitedefaultmidpunct}
{\mcitedefaultendpunct}{\mcitedefaultseppunct}\relax
\EndOfBibitem
\bibitem{LHCb-TDR-016}
{LHCb collaboration}, \ifthenelse{\boolean{articletitles}}{\emph{{LHCb Trigger
  and Online Upgrade Technical Design Report}}, }{}
  \href{http://cdsweb.cern.ch/search?p=CERN-LHCC-2014-016&f=reportnumber&action_search=Search&c=LHCb+Reports}
  {CERN-LHCC-2014-016}, 2018\relax
\mciteBstWouldAddEndPuncttrue
\mciteSetBstMidEndSepPunct{\mcitedefaultmidpunct}
{\mcitedefaultendpunct}{\mcitedefaultseppunct}\relax
\EndOfBibitem
\bibitem{LHCb-DP-2019-001}
R.~Aaij {\em et~al.}, \ifthenelse{\boolean{articletitles}}{\emph{{Performance
  of the LHCb trigger and full real-time reconstruction in Run 2 of the LHC}},
  }{}\href{https://doi.org/10.1088/1748-0221/14/04/P04013}{JINST \textbf{14}
  (2019) P04013}, \href{http://arxiv.org/abs/1812.10790}{{\normalfont\ttfamily
  arXiv:1812.10790}}\relax
\mciteBstWouldAddEndPuncttrue
\mciteSetBstMidEndSepPunct{\mcitedefaultmidpunct}
{\mcitedefaultendpunct}{\mcitedefaultseppunct}\relax
\EndOfBibitem
\bibitem{LHCb-TDR-013}
{LHCb collaboration}, \ifthenelse{\boolean{articletitles}}{\emph{{LHCb VELO
  Upgrade Technical Design Report}}, }{}
  \href{http://cdsweb.cern.ch/search?p=CERN-LHCC-2013-021&f=reportnumber&action_search=Search&c=LHCb+Reports}
  {CERN-LHCC-2013-021}, 2013\relax
\mciteBstWouldAddEndPuncttrue
\mciteSetBstMidEndSepPunct{\mcitedefaultmidpunct}
{\mcitedefaultendpunct}{\mcitedefaultseppunct}\relax
\EndOfBibitem
\bibitem{LHCb-DP-2013-002}
LHCb collaboration, R.~Aaij {\em et~al.},
  \ifthenelse{\boolean{articletitles}}{\emph{{Measurement of the track
  reconstruction efficiency at LHCb}},
  }{}\href{https://doi.org/10.1088/1748-0221/10/02/P02007}{JINST \textbf{10}
  (2015) P02007}, \href{http://arxiv.org/abs/1408.1251}{{\normalfont\ttfamily
  arXiv:1408.1251}}\relax
\mciteBstWouldAddEndPuncttrue
\mciteSetBstMidEndSepPunct{\mcitedefaultmidpunct}
{\mcitedefaultendpunct}{\mcitedefaultseppunct}\relax
\EndOfBibitem
\bibitem{Quagliani:2296404}
R.~Quagliani, \ifthenelse{\boolean{articletitles}}{\emph{{Study of double charm
  B decays with the LHCb experiment at CERN and track reconstruction for the
  LHCb upgrade}}, }{} 2017.
\newblock Ph.D. thesis, presented Oct 6$^{th}$ 2017\relax
\mciteBstWouldAddEndPuncttrue
\mciteSetBstMidEndSepPunct{\mcitedefaultmidpunct}
{\mcitedefaultendpunct}{\mcitedefaultseppunct}\relax
\EndOfBibitem
\bibitem{LHCb-DP-2019-003}
LHCb collaboration, R.~Aaij {\em et~al.},
  \ifthenelse{\boolean{articletitles}}{\emph{{Measurement of the electron
  reconstruction efficiency at LHCb}},
  }{}\href{https://doi.org/10.1088/1748-0221/14/11/P11023}{JINST \textbf{14}
  (2019) P11023}, \href{http://arxiv.org/abs/1909.02957}{{\normalfont\ttfamily
  arXiv:1909.02957}}\relax
\mciteBstWouldAddEndPuncttrue
\mciteSetBstMidEndSepPunct{\mcitedefaultmidpunct}
{\mcitedefaultendpunct}{\mcitedefaultseppunct}\relax
\EndOfBibitem
\bibitem{hough1962method}
P.~V. Hough, \ifthenelse{\boolean{articletitles}}{\emph{Method and means for
  recognizing complex patterns}, }{} 1962.
\newblock US Patent 3,069,654\relax
\mciteBstWouldAddEndPuncttrue
\mciteSetBstMidEndSepPunct{\mcitedefaultmidpunct}
{\mcitedefaultendpunct}{\mcitedefaultseppunct}\relax
\EndOfBibitem
\bibitem{LHCb-TDR-021}
LHCb collaboration, \ifthenelse{\boolean{articletitles}}{\emph{{LHCb Upgrade
  GPU High Level Trigger Technical Design Report}}, }{}
  \href{http://cdsweb.cern.ch/search?p=CERN-LHCC-2020-006&f=reportnumber&action_search=Search&c=LHCb+Reports}
  {CERN-LHCC-2020-006}, 2020\relax
\mciteBstWouldAddEndPuncttrue
\mciteSetBstMidEndSepPunct{\mcitedefaultmidpunct}
{\mcitedefaultendpunct}{\mcitedefaultseppunct}\relax
\EndOfBibitem
\end{mcitethebibliography}

\end{document}